\documentclass{elsart}

\usepackage{epsfig}

\usepackage{amssymb}

\begin{document}

\begin{frontmatter}



\title{Comparison of muon and neutrino times from decays of mesons in the atmosphere}


\author[label1]{Teresa Montaruli},
\author[label2]{Francesco Ronga}

\address[label1]{D\'epartement de physique nucl\'eaire et corpusculaire, Universit\'e de Gen\`eve, CH-1211}
\address[label2]{INFN Laboratori Nazionali di Frascati Frascati I-00044}
\corauth[cor]{Corresponding author; email: teresa.montaruli@unige.ch}

\begin{abstract}
We propose a test to compare the speed of muons and muon neutrinos produced in the same bundles
in the atmosphere. By comparing the arrival times of primary muons and muons induced by neutrinos, experiments such as Super-Kamiokande, IceCube and ANTARES
could verify recent hints on the possibility that neutrinos have a velocity larger than the speed of light.
For smaller detectors such as MINOS and OPERA, where it may be challenging to recognize downgoing muons induced by neutrinos inside the active volume,
another measurement may be possible thanks to the ns-time resolution. The arrival times of muons in a bundle produced by one cosmic ray interaction in the atmosphere
should return a value compatible with the time resolution, while if neutrinos have larger velocity than muons some of the muons may arrive 
slightly earlier in a bundle of particles than others.

\end{abstract}

\end{frontmatter}

\section{Introduction}
\label{sec1}
With long baseline neutrino experiments it is possible to measure the neutrino velocity by the time difference between neutrino detection and production.
Recently the OPERA neutrino experiment  on the 730 km CERN-Gran Sasso neutrino beam has measured a difference of $60.7\pm6.9(stat)\pm7.4(sys)$ ns 
with respect to the time computed assuming that muon neutrinos are traveling with the speed of light observed by OPERA. 
There is no apparent energy dependence of this time difference between $\sim 10 - 40$ GeV, the energy spread of the CNGS beam. This result, if confirmed, would imply that neutrino are tachyons with speed above $v$ the speed of light by $\frac{v-c}{c} = 2.48 \pm 0.28(stat)\pm 0.30(sys) \times 10^{-5}$ \cite{OPERA:2011zb}.

This result is in agreement with the result of the MINOS experiment that applied a similar technique but with a much larger error \cite{Adamson:2007gu}. The OPERA result is however  in disagreement with the observation of anti-electron neutrinos of energies of about 10 MeV from the famous supernova SN1987A providing an upper limit of of $ |{v-c}|/c < 2\times 10^{-9} $ \cite{sn1987}. If the effect is real, then this could hint to a possible energy dependent effect or other yet unknown flavor dependent effect. 

The interaction of a primary cosmic ray  with the atmosphere produces a cascade with many kind of particles, and in particular neutrinos and muons. Muon neutrinos and muons are produced  mainly via the decay of charged pions and kaons produced in the primary cosmic ray interactions and above about 10 TeV they can come also from prompt decays of charmed hadrons. This last component has not yet been observed. The  neutrinos in this cascade, if the OPERA result is correct, should arrive before the muons from the same parent decay with a time delay that should change according to the neutrino path length that depends on its zenith angle $\theta$. In underground detectors muon neutrinos are detected looking for induced muons produced by neutrino charged current interactions in the rock, and the ice around or inside the instrumented region. Hence, a time spread should be observed between the muons produced by the pion or kaon decay directly and the muon produced by neutrino interactions.

The  path length from the meson decay point is a few tens of kilometers for vertical neutrinos  and up to $\sim 300$~km for near horizontal neutrinos. 
Assuming that the time difference observed in OPERA is not energy dependent and it is 60 ns, nearly horizontal neutrinos should arrive up to 28 nsec before the other secondaries. In Ref.~\cite{Stanev} a table of average production heights neutrinos in the atmosphere has been reported. The typical production height for neutrinos of energy above 20 GeV can be 17.6 km at the vertical, 94.9 km at $\cos\theta = 0.25$ and 335.7 km at $\cos\theta = 0.05$, which would correspond to
1.4, 78 and 27.6~ns. The typical time resolutions of photomultipliers (PMTs), including the associated electronics, is between $\Delta t \sim 1-3$~ns \cite{timing_antares,icecubePMT,icecube_first}.
Typical tracks are reconstructed using a number of PMT time hits $N$ and the time uncertainty associated to a muon track depends such number as $\Delta t/\sqrt N$.

There are already limits to tachions in cosmic rays. The limits are  obtained looking for example to signals before the main front of the electromagetic shower. But this kind of searches stopped some time ago and the last particle data book  review related to this tachion searches is the one of 1994 \cite{montanet}. The  limits obtained are of small interest in the framework of the OPERA result. However, if neutrinos are tachions, it is likely that other kind of tachions could exist and this search in very high energy cosmic rays could have a new interest. 
 
The MACRO experiment has done several searches for possible anomalies of the time differences between  muons
\cite{Ahlen:1991yh,Becherini:2005xg}. The searches  were dedicated mainly to study time differences of the order of a few msec or more, but Ref.~\cite{Ahlen:1991yh} contains also the study of time differences at the ns-level.  The results of this study was that the largest time differences between two muons was 18 ns and that the time distribution was in agreement with the predictions.  The statistics was limited to 35832 tracks in events with two or more tracks. In 1992 none was thinking to  tachionic neutrinos and therefore there was no estimate of the number of tracks due to down-going neutrino together with  a primary muons. In Ref~\cite{Scappa} this study was extended to  about 140000 tracks of multi muon events, corresponding to about $4\%$ of the total MACRO statistics. The time distribution was in agreement with the predictions.

Currently the largest neutrino detector in operation is the IceCube detector at the South Pole. We will discuss in the following paragraphs the possibility to measures the time delay between the muon produced by the neutrino interaction in the ice  and the muon produced directly in the atmosphere using IceCube and the compact detector inside IceCube called DeepCore. DeepCore, vetoed by the surrounding IceCube strings, has the possibility to separate events produced internally by neutrinos and the muons directly produced in the decays in the atmosphere. 
The real analysis is made difficult by the scarce density of photodetectors that prevents a reliable tracking of multiple muons in one event but the veto possibility offered by the PMTs surrounding DeepCore and the large dimensions of IceCube may make this measurement interesting. In fact the tracking can be applied to clusters of hits recognized by
`hit cleaning' algorithms that can be applied to recognize different topologies in the same time frame of the data acquisition.
ANTARES has better time resolution for muon tracks than IceCube due to the smaller amount of light scattering. A similar analysis may be attempted there though the
identification of the neutrino may be more tricky in the absence of a veto. Super-Kamiokande can measure internal neutrino events and has good pattern recognition capabilities
as well as OPERA and MINOS that have also excellent time of flight measurement capabilities. Even if these detectors have limited dimension and may not
be able to identify internal downgoing neutrinos, they have good tracking capabilities and may perform a measurement of muon bundle timing. All muons should arrive at the same time compatibly with the time resolution, but muons induced by neutrinos could arrive earlier if they are tachions.

\section{The semi-analytical atmospheric neutrino flux calculation}

We have calculated the atmospheric muon neutrino flux using the tables provided by the japanese group in 
\cite{hondatables}. Their calculation is described in \cite{honda2006}.
We have fit these tables for muon neutrinos and anti-neutrinos with a  5 degree-polynomial
function of $(x,y)=(\cos\theta, log_{10}E)$, where E and $\theta$ are the energy and nadir angle of the neutrino respectively:
\begin{equation}
log_{10}\left(\frac{dN_{\nu}}{d E}\right) =\sum_{i= 1}^{5} p_{ix} x^{i} + p_{0} +  \sum_{i=1}^{5} p_{iy} y^{i}
+ \sum_{i=1}^{4} p_{ixy} x^{i} y^{5-i}
\label{eq:poly}
\end{equation}

The parameters of the fit are given in Tab.~\ref{tab:polynumu}.
\begin{table}
\begin{center}
\begin{tabular}{|ccccc|}
\hline
 Ref. &  $p_{5x}$ & $p_{4x}$& $p_{3x}$ & $p_{2x}$ \\
 & $p_{1x}$  & $p_{0}$ &$p_{5y}$ & $p_{4y}$\\
 &  $p_{3y}$ &$p_{2y}$&$p_{1y}$&$p_{1xy}$ \\
 &$p_{2xy}$ & $p_{3xy}$ & $p_{4xy}$& \\ \hline
HKKM2006 $\nu_{\mu}$ & $-8.79638 \times 10^{-1}$ & $-1.17432$ &
$-9.39143 \times 10^{-1}$ & $1.34865 \times 10^{-1}$ \\
& $6.88957 \times 10^{-1}$& $-1.43770$ & $7.25078 \times 10^{-4}$ & $-8.69003 \times 10^{-3}$ \\
& $3.39606 \times 10^{-2}$ & $-1.04246 \times 10^{-1}$ & $-2.76920$ &
$2.65441 \times 10^{-3}$ \\
& $-1.07521 \times 10^{-2}$ & $-1.26859 \times 10^{-1}$ &
$-4.20873  \times 10^{-1}$ & \\ \hline
HKKM2006 $\bar{\nu}_{\mu}$ & $-1.14682$ & $-1.68745$ & 
$-1.10493$ & $4.10708 \times 10^{-1}$ \\
& $9.74675 \times 10^{-1}$& $-1.39129$ & $7.54875 \times 10^{-4}$ & $-9.12622 \times 10^{-3}$ \\
& $3.56607 \times 10^{-2}$ & $-1.12264 \times 10^{-1}$ & $-2.81577$ &
$1.37765 \times 10^{-3}$ \\
& $-1.88374 \times 10^{-2}$ &$-1.52999 \times 10^{-1}$&
$-4.52546 \times 10^{-1}$ & \\ \hline
\end{tabular}
\caption{\label{tab:polynumu} Parameters for the 5 degree polynomial in x = $\cos\theta$ and y = $log_{10} E$ as described by eq.~\ref{eq:poly} for the calculation in Ref.~\protect\cite{honda2006,hondatables} for muon neutrinos and anti-neutrinos.}
\end{center}
\end{table}

Between energies of $\sim 10-100$~GeV the flux of atmospheric neutrinos is not a single power law
but above about 500 GeV the flux can be described by the analytical function that we reproduce here from Ref.~\cite{tombook}:
\begin{equation}
\frac{dN_{\nu_{\mu}}}{d lnE} = E^{\gamma} \left[ \frac{A_{\nu}}{1+B_{p}E \cos\theta^{*}/\epsilon_{\pi}} + \frac{B_{\nu}}{1+B_{k}E \cos\theta^{*}/\epsilon_{k}} \right]
\label{eq:numu}
\end{equation}
where $\cos\theta^{*}$ is equal to $\cos\theta$ for $\cos\theta < 0.3$ and for horizontal directions ($\cos\theta > 0.3$) has a larger value that accounts for the curvature of the Earth.
Notice that we consider the neutrino fluxes at these energies up-down symmetric.
Above 500 GeV, muon decay does not contribute to the neutrino flux at these energies and 
scaling of hadron-nucleus interactions starts to dominate.
Originally the neutrino flux formulas given above were derived assuming a flat atmosphere, and are therefore only valid up to nadir angles of $\lesssim 70^{\circ}$. In order to describe the full range of zenith angles, $0^{\circ}-90^{\circ}$, the $\cos(\theta)$ dependence of the $\pi$ and $K$ critical energies valid at zenith angles below $60^{\circ}$ can be replaced with a $\cos(\theta^*)$ dependence as in \cite{volkovaZ}. In this work a convenient parameterization of this substitution is given:

\begin{equation}
\cos\theta^* = \sqrt{\frac{\cos^2\theta + p_{0}^2 + p_1 \cdot \cos^{p_{2}} \theta + p_3\cdot\cos^{p_{4}}\theta } {1+p_0^2+p_1+p_3}}
\label{eq:costheta}
\end{equation}
as given in \protect\cite{dima} and parameters are from the same paper (see eq.~\ref{tab:cos}).
For muon neutrinos the low energy function described in eq.~\ref{eq:poly} and the high energy functions have been connected at 500 GeV. The parameters that have been tuned based on the Monte Carlo calculation in the region 500 GeV - 100 TeV are the normalizations of the pion and kaon terms, respectively $A_{\nu}$ and $B_{\nu}$ and the spectral index $\gamma$.
The parameters of the function in eq.~\ref{eq:numu} are given in Tab.~\ref{tab:numu} and differ for neutrinos and
antineutrinos. 

\begin{table}
\begin{center}
\begin{tabular}{|ccccc|}
\hline
$p_0$ & $p_1$ & $p_2$ & $p_3$ & $p_4$\\                      
\hline
$1.02573 \times 10^{-1}$ & $-6.82870 \times 10^{-2}$ & $9.58633 \times 10^{-1}$ & $4.07253 \times 10^{-2}$ &
$8.17285 \times 10^{-1}$ 
 \\  \hline
\end{tabular}
\caption{\label{tab:cos} Parameters for the $\cos\theta^{*}$ function described in eq.~\ref{eq:costheta}.
}
\end{center}
\end{table}

\begin{table}
\begin{center}
\begin{tabular}{|cccccccc|}
 \hline
 Ref. &  $\gamma$ & $A_{\nu}$& 
 $B_{p}$ &$\epsilon_{\pi}$  & $B_{\nu}$ &$B_{k}$ & $\epsilon_{k}$ \\
                                     &                      &  (cm$^{-2}$ s$^{-1}$ sr$^{-1}$)&  & (TeV)& (cm$^{-2}$ s$^{-1}$ sr$^{-1}$)& 
                        & (TeV)  \\
\hline
HKKM2006  $\nu_{\mu}$&  -1.7 & $8.90369 \times 10^{-8}$& 2.77 & 0.115& $2.649640 \times 10^{-8}$ & 1.18& 0.85  \\  
HKKM2006  $\bar{\nu}_{\mu}$ & -1.7 & $7.85269 \times 10^{-8}$& 2.77 & 0.115& $1.36649 \times 10^{-8}$ & 1.18& 0.85  \\  \hline
\end{tabular}
\caption{\label{tab:numu} Parameters for the high energy functions in eq.~\protect\ref{eq:numu} for $\nu_{\mu}$ and 
$\bar{\nu}_{\mu}$ fluxes
EdN/dE in units cm$^{-2}$ s$^{-1}$ sr$^{-1}$, with the energy in TeV for HKKM2006 \protect\cite{honda2006} as derived fitting the tables in the energy region 500~GeV-100~TeV.}
\end{center}
\end{table}

\section{Calculation of the muon number of events accompanying neutrinos underground}

The IceCube detector was recently completed and complemented by the DeepCore array dedicated to enhancing IceCube capabilities 
for neutrino detection in the 10-100~GeV energy range. The filter to select internal events vetoed by external layer of PMTs
is described in \cite{DeepCore}, where the effective area is given. 

In Ref.~\cite{Schoenert} the calculation of the efficiency of the veto has been applied focused to neutrino astrophysics studies: neutrinos
directly produced in sources could be distinguished by neutrinos produced in the atmosphere by vetoing events where also a muon
is detected together with the muon-induced neutrino inside DeepCore. We reverse the problem here: we want to keep both the 
external muon and the internal muon-induced neutrino to measure the difference in time described in the Introduction.
Hence, we calculate the muon event rate that can be detected in IceCube and that would accompany a neutrino that can induce a muon detectable in DeepCore or
in the rest of IceCube provided that the track is not starting outside IceCube.
We consider pion and kaon decays into muon neutrinos and muons and ignore other minor contributions, for instance of three-body kaon decays. The kinematics of this two body decay is described in Ref.s~\cite{tombook,Schoenert}.
We assume that the standard kinematics can be used for the superluminal neutrinos. This is in general not true, but this assumption is justified experimentally from the agreement of the experimental data of the two body pion and kaon decays with the standard kinematics.

 In this paper the separation between the muon and the muon induced neutrino is calculated and it is about 1m (0.1 m) for neutrino energies above 1 TeV (10 TeV) for pion primaries and 10 m (1 m) for kaons. These values may dependent on the zenith angle.
Tracks that are separated by few meters in IceCube are a challenge for reconstructions but if the neutrino event may be recognized in the inner denser core of 
IceCue and the muon entrance in the detector is detected in the external part the measurement is possible.
The fraction of primary energy taken by muons and muon neutrinos depends on the mass of the primary. 
The distribution of neutrino energy in the decay in the atmosphere are flat over the kinematically allowed regions between $0 \le E_{\nu} \le E_i (1- r_i)$, where $E_i$ is the primary pion or kaon energy and
$r_i = m_{\mu}^2/m_{i}^2 = 0.573$ for $\pi^{\pm}$ and 0.046 for $K^{\pm}$. Hence the remaining energy goes to the muon and it is limited by the condition 
$E_{\mu} \ge E_{\nu} \times \frac{r_i}{1-r_i}$ \cite{Schoenert}.
For kaon decays this fraction is much smaller than for pion decays. The fraction of neutrinos from kaons and pions as a function of energy is mildly dependent 
on the angle up to $\theta = 60^o$. We took these values from Ref.~\cite{gaisserhonda}.

The minimum muon energy from the decay corresponding to a neutrino energy $E_{\nu}$ has to be enough to be detected underground. 
The minimum muon energy to penetrate at a depth $X$ (in km w. e.) was taken from Ref.~\cite{dima} and differs a little from what quoted in Ref.~\cite{Schoenert}: $E_{\mu , min}  = 0.57 \times [(\exp(X/2.1)-1]$~TeV, where we take $X = h/cos\theta^*$, $h = 1.95$ km w.e. is the depth of the middle point in IceCube and $\cos\theta^*$ is given in eq.~\ref{eq:costheta} and accounts for the Earth curvature. This expression for the minimum muon energy comes from an average estimate of energy losses
of muons in ice including stochastic effects from montecarlo: $- < dE/dX>  = a + bE_{\mu}$, where a = 0.268 GeV/m w.e. and $b = 0.470 \times 10^{-3}$ m w.e.$^{-1}$. Integrating  $dX/dE$, the inverse f the energy loss, over energy one obtains the well known formula $E_{\mu ,min} = a/b \times (e^{bX} -1)$ \cite{tombook}. 

\begin{figure}
\begin{tabular}{cc}
\epsfig{file = 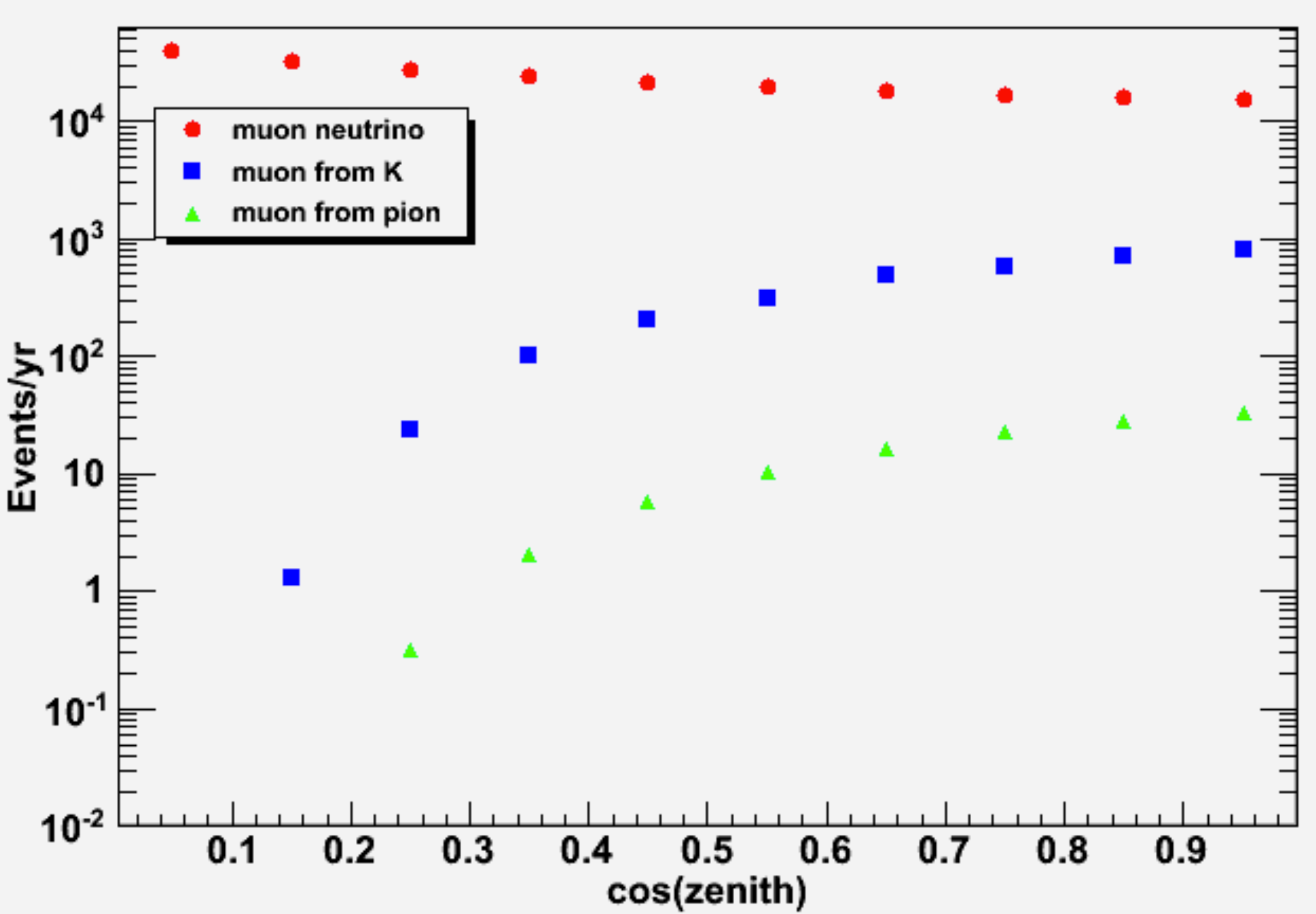, height=5cm} &
\epsfig{file = 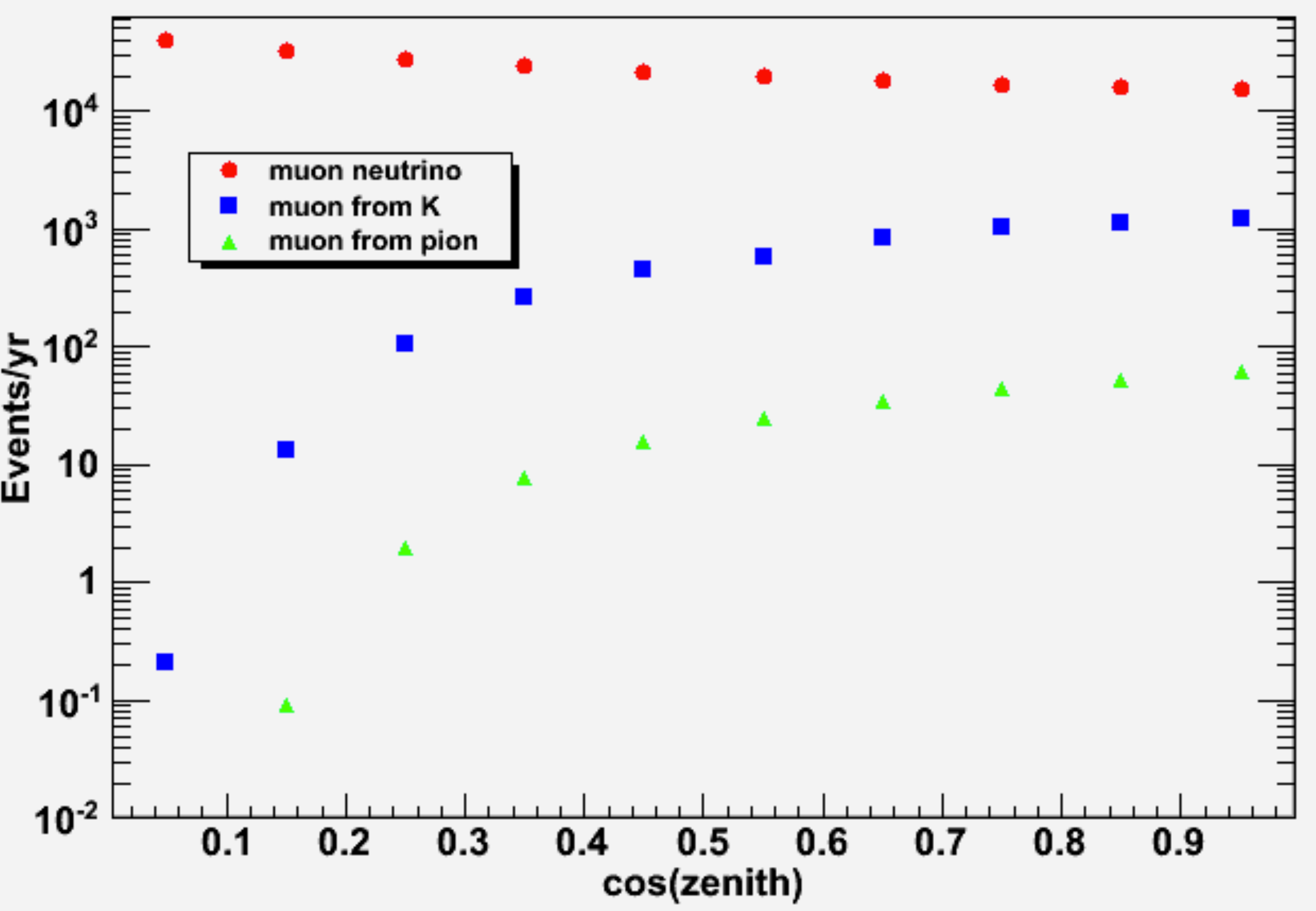, height=5cm} 
\end{tabular}
\caption{\label{fig1} The upper curve is the number of downgoing muon neutrino events per year in 86 strings of IceCube as a function of
angle, the other two curves are the number of accompanying muons from pion and from kaon decays. On the left we have
assumed the depth of the core of IceCube of 1.95 km w.e. and on the right the top of the array depth of 1.5 km we. In the first case the total number of muon events is about 3231
and in the second it is 5851 muon events/yr.
}
\end{figure}

We show in Fig.~\ref{fig1} the number of atmospheric muon neutrino events detected in IceCube as a function of $\cos\theta$ where we consider an energy dependent effective area including the effect of DeepCore \cite{DeepCore} and extrapolate to 86 strings the effective area given in Ref.~\cite{ic40} that includes the efficiency for muon detection. We neglect dependencies on the angle
of the effective area. We show also the number of accompanying muons from kaons and pions. 
The total number of events is about 3231 per year. The numbers as a function of $\cos\theta$ are given in Tab.~\ref{tab:result} as well.
The measurement of the time difference between muons as a function of the angle would test the energy dependence of the effect seen in OPERA. As a matter of fact the horizontal muons need to have higher energies than vertical ones.
 Since the veto efficiency is about 40\% at 1 TeV we can expect that in order to recognize the neutrinos these 
 numbers will be further reduced but in a way that requires a full simulation for a correct estimate.

 For detectors that have good tracking capabilities of multiple muons in a bundle or for muons in a bundle that have high enough separation, 
the efficiency for this search can be improved not applying any veto requirement to identify muon induced neutrinos. The times of arrival of muons in a bundle 
can be measured.
Due to the high energies involved, we expect sub-nanosecond delays between muons, even if produced by different decays of mesons in the shower.  The average distance between muons was measured by the MACRO experiment  \cite{decoherence} at the Gran Sasso laboratory depth and  changes from 7 meter for vertical incidence up to 19 meters for $cos\theta=0.55.$ 
This average separation is small compared to the IceCube dimensions, nonetheless muons with higher separations exist and can be detected still with 
significant statistics given the large dimensions (see eg. \cite{gerhardt}).

Very roughly the probability  that a muon produced by a neutrino is accompanied by another muon from another meson that decays in the same shower
is of the order of $5-10\%$  (the typical ratio between double and single muons in a muon bundle). This is true if the neutrino energy is of the order of  $E_{\mu , min}$.
In Ref.~\cite{spurio} it is estimated that the double muons
 are about 10\% (5\%) of single muons at  3 km w.e. of depth for primary directions of $\theta = 0^o$ ($\theta = 70^o$). 
Small detectors  like MINOS, OPERA (and in the past MACRO) have limited  possibilities to identify downgoing muons from neutrino interactions.  But, if the time resolution is very good, the study of  the time difference of the events with two or more muons could be interesting even without the neutrino signature. 


\begin{table}
\begin{center}
\begin{tabular}{|ccccc|}
 \hline
 $\cos\theta$ &  Neutrinos & Muons from pions & Muons from kaons& Total Muons\\
 \hline
0.05& 39677.3  & 0  & 0  & 0\\ \hline  
0.15 & 32852.3 & 1.3& 0 & 1.3 \\  \hline
0.25&  27680.8 & 23.8 & 0.3 & 24.1 \\ \hline  
0.35 & 24052.7 & 99.9 &  2.1& 102.0\\  \hline
0.45&  21474.8 & 204.5 & 5.7 & 210.3\\ \hline  
0.55 & 19570.1  & 307.7 & 10.2  & 317.9\\ \hline
0.65& 18091.0& 495.7 & 16.1 & 511.9\\ \hline  
0.75 & 16903.9 & 586.5 &23.1 &609.6 \\  \hline
0.85& 15960.9 & 705.4 & 27.5 & 732.8  \\ \hline  
0.95 & 15336.9 & 805.9 & 32.9  & 838.8\\  \hline
Total & 231,600 & 3230.8  & 117.9& 3348.7\\ \hline 
\end{tabular}
\caption{\label{tab:result} Number of atmospheric muon neutrinos \protect\cite{honda2006} and corresponding detectable muons per year accompanying  a neutrino in IceCube per year at a depth of 1.95 kmwe (the core of IceCube). The total in the last  row is integrated over the downgoing hemisphere.}
\end{center}
\end{table}

\section{Conclusions}

We have made an approximate calculation of the number of muon events accompanying neutrino events in a detector like IceCube.
The determination of the arrival time of the muons and the difference of the arrival of the accompanying neutrinos
could be used to verify in the atmospheric neutrino sector the OPERA observation. Also the time differences between muons in a bundle may indicate
that a part of them may have arrived a little earlier. This would be the case for the muons induced by neutrinos.

\section*{Acknowledgements}
TM would like to thank T.K.~Gaisser for the idea on how to extrapolate at high energy the neutrino fluxes and
Dima Chirkin for pointing out the $\cos\theta$ effective function in eq.~\ref{eq:costheta} \cite{dima} and providing the minimum energy of muons at a certain depth with parameters 
from MMC \cite{dima}.


\end{document}